\documentclass[fleqn,usenatbib]{mnras}
\usepackage{newtxtext,newtxmath}
\usepackage[T1]{fontenc}
\usepackage{ae,aecompl}

\usepackage{graphicx} 
\usepackage{amsmath,amssymb,times}
\usepackage{longtable}
\usepackage{epstopdf}
\usepackage[usenames,dvipsnames]{color}
\usepackage{float}
\usepackage[export]{adjustbox}
\usepackage{ulem}
\usepackage[export]{adjustbox}
\usepackage{hyperref}
\usepackage{pdflscape}
\usepackage{siunitx}

\hypersetup{
  colorlinks= true, 
  urlcolor  = Blue, 
  filecolor=black,
  runcolor=blue,
  linkcolor = red, 
  citecolor = blue 
} 
\makeatletter

\newcommand{\OIII}{O\,{\scriptsize III}}
\newcommand{\OII}{O\,{\scriptsize II}}

\newcommand{\NII}{N\,{\scriptsize II}}
\newcommand{\CIII}{C\,{\scriptsize III}}

\newcommand{\NV}{N\,{\scriptsize V}}
\newcommand{\CIV}{C\,{\scriptsize IV}}
\newcommand{\HeII}{He\,{\scriptsize II}}
\newcommand{\HII}{H\,{\scriptsize II}}
\newcommand\pegase{P\'egase3}
\makeatother
\makeatletter
\def\@to{to}
\makeatother

\title[The \textit{Spitzer}/IRAC Colours of 7$\leqslant z \leqslant$9 Galaxies]{Interpreting the \textit{Spitzer}/IRAC Colours of 7$\leqslant z \leqslant$9 Galaxies: Distinguishing Between Line Emission and Starlight Using ALMA}

\author[Roberts-Borsani et al.]{G.~W. Roberts-Borsani$^{1,2}$\thanks{E-mail: guidorb@astro.ucla.edu}, R.~S. Ellis$^{2}$ and N. Laporte$^{3,4,2}$
\\
$^{1}$Department of Physics and Astronomy, University of California, Los Angeles, 430 Portola Plaza, Los Angeles, CA 90095, USA \\
$^{2}$Department of Physics and Astronomy, University College London, Gower Street, London WC1E 6BT, UK \\
$^{3}$Kavli Institute for Cosmology, University of Cambridge, Madingley Road, Cambridge CB3 0HA, UK \\
$^{4}$Cavendish Laboratory, University of Cambridge, 19 JJ Thomson Avenue, Cambridge CB3 0HE, UK}

\date{Accepted XXX. Received YYY; in original form ZZZ}
\pubyear{2019}

\begin{document}
\label{firstpage}
\pagerange{\pageref{firstpage}--\pageref{lastpage}}
\maketitle

\begin{abstract}
Prior to the launch of JWST, \textit{Spitzer}/IRAC photometry offers the only means of studying the rest-frame optical properties of $z>$7 galaxies. Many such high redshift galaxies display a red [3.6]$-$[4.5] micron colour, often referred to as the ``IRAC excess'', which has conventionally been interpreted as arising from intense [\OIII]+H$\beta$ emission within the [4.5] micron bandpass. An appealing aspect of this interpretation is similarly intense line emission seen in star-forming galaxies at lower redshift as well as the redshift-dependent behaviour of the IRAC colours beyond $z\sim$7 modelled as the various nebular lines move through the two bandpasses. In this paper we demonstrate that, given the photometric uncertainties, established stellar populations with Balmer (4000 \AA\, rest-frame) breaks, such as those inferred at $z>$9 where line emission does not contaminate the IRAC bands, can equally well explain the redshift-dependent behaviour of the IRAC colours in 7$\lesssim z\lesssim$9 galaxies. We discuss possible ways of distinguishing between the two hypotheses using ALMA measures of [\OIII] $\lambdaup$88 micron and dust continuum fluxes. Prior to further studies with JWST, we show that the distinction is important in determining the assembly history of galaxies in the first 500 Myr.
\end{abstract}


\section{Introduction}
The last few years has seen impressive progress in studies of galaxies in the so-called ``reionisation era'' corresponding to the redshift interval $7<z<10$. However, the number of spectroscopically-confirmed examples remains limited and much has been deduced from spectral energy distributions (SEDs) of photometric samples. In addition to demographic studies based on star formation rate densities \citep{oesch14,mcleod16} and luminosity functions \citep{atek15,bouwens15}, a key area of interest is studies of the gaseous and stellar properties of early systems. The latter topic is central to understand both the ionising capability of early galaxies as well as the age of their stellar populations (for a recent review see \citealt{stark16}).

Although much of the progress has been made using photometric samples based on Hubble imaging, both in deep fields \citep{grogin11,koekemoer11,ellis13} and through lensing clusters \citep{bradley14,lotz17,salmon18,coe19}, the \textit{Spitzer Space Telescope} has made a key contribution since, at $z\gtrsim$5, the two bandpasses at 3.6 and 4.5 $\mu$m sample the rest-frame optical. At redshifts of $z\gtrsim$6.6-6.8, it is claimed that the redshift-dependent trend of the IRAC colours is consistent with intense nebular emission lines shifting through the bandpasses \citep{labbe13,smit15,rb16}, although the precision of this exercise is dependent mostly on samples with only photometric redshifts. With this in mind, the surprising spectroscopic confirmation with Lyman-$\alpha$ \citep{oesch15,zitrin15,stark17} of the 4 brightest $z>7$ galaxies in the CANDELS survey selected to display red \textit{Spitzer}/IRAC 3.6 $\mu$m$-$ 4.5 $\mu$m colours (and hence an ``IRAC excess'') of $>0.5$ mag \citep{rb16}, reinforced the hypothesis that the IRAC excess arises from intense [\OIII] $\lambdaup\lambdaup$ 4959,5007 \AA\ plus H$\beta$ emission within the 4.5 $\mu$m band. To explain the IRAC colours, the rest-frame equivalent widths (EWs) of [\OIII]+H$\beta$ must be of order 500 \AA.

Although it will not be possible to confirm this suggestion with direct spectroscopy until the launch of JWST, the so-called ``[\OIII] hypothesis" has been widely accepted for several reasons. Firstly, at lower redshift $z\sim$6.6-6.8 where [\OIII] passes through the 3.6 $\mu$m bandpass, the required blue 3.6 $\mu$m$-$4.5 $\mu$m colour is seen for a sample of galaxies, several of which are now spectroscopically-confirmed with ALMA (\citealt{smit18}; see also \citealt{sobral15},\citealt{pentericci16},\citealt{matthee17}). Finally, as a proof of concept, galaxies whose rest-frame [\OIII] emission exceed EW$\simeq$1000 \AA\ , while difficult to reproduce via modelling except in very young star-forming systems, have been studied at $z\simeq$2 \citep{maseda14}.

A more recent development has been the location of IRAC-excess galaxies whose photometric redshifts lie at $z>$9; at these redshifts [\OIII] $\lambdaup$5007 \AA\ - the strongest contaminating emission line - is shifted beyond both IRAC filters. Although both H$\beta$ and [\OIII] $\lambdaup$4959 \AA\ remain in the filter until $z\sim$9.3, their relatively low strengths make it difficult to reproduce strong IRAC-excesses. Thus far only one system, MACS1149-JD1, hereafter JD1, has been spectroscopically-confirmed at $z$=9.11 \citep{hashimoto18}. Analysis of its SED attributes the IRAC excess to the Balmer break at 4000 \AA\ consistent with a mature $\sim$200-300 Myr old stellar population providing a first tantalising glimpse of ``cosmic dawn'' at $z\simeq$15$\pm$3. This discovery raises the question of the extent to which the IRAC excess seen in galaxies at 7$<z<$9 might also, in part, be due to a similar Balmer break. The distinction is important since it would imply many luminous $z\simeq$7-9 galaxies may have older stellar populations and larger stellar masses than previously thought, with interesting consequences for the presence of earlier star formation. An additional issue is whether JD1 is representative of the galaxy population at $z\simeq7-9$ (see \citealt{katz19}).

The goal of the present paper is to explore the extent to which an IRAC excess and its redshift-dependent trend might be due, in part, to starlight rather than solely intense [\OIII] line emission. A plan of the paper follows. In \S2 we examine predicted IRAC colours in the context of both hypotheses, using carefully-chosen template galaxies as well as contemporary stellar population models that incorporate nebular line emission. In \S3, we turn to what data might be needed to distinguish between the two hypotheses. Prior to spectroscopy with JWST, we consider the flux of [\OIII] $\lambdaup$88 $\mu$m which is accessible with ALMA and examine the IRAC colour for those spectroscopically-confirmed galaxies for which [\OIII] $\lambdaup$88 $\mu$m fluxes are available. In \S4 we discuss our results and the implications on the early assembly of galaxies. Throughout this paper we refer to the \textit{HST} F160W and \textit{Spitzer}/IRAC 3.6 and 4.5 micron bands as \textit{H}$_{160}$, [3.6] and [4.5], respectively, for simplicity. We also assume \textit{H}$_{0}$=70 km/s/Mpc, $\Omega_{m}=$ 0.3, and $\Omega_{\wedge}=$0.7. All magnitudes are in the AB system \citep{oke83}.

\section{Modelling the IRAC Colours of 7$<z<$9 Galaxies}
\label{sec:iracmodelling}
In exploring the redshift-dependent behaviour of the IRAC 3.6 and 4.5 $\mu$m colours, under the hypotheses of contributions from [\OIII] line emission or a Balmer break due to a more mature stellar population, we begin by selecting two template SEDs fitted to actual data for spectroscopically-confirmed $z>$7 galaxies. For the case of intense [\OIII] emission, we use a spectroscopic template fitted to the ground-based and \textit{HST}/\textit{Spitzer} photometry of EGSY8p7 at $z$=8.68 \citep{zitrin15}, one of the four IRAC-excess bright sources first identified in the CANDELS survey \citep{rb16}. The template and \textit{HST}/\textit{Spitzer} photometry used here are taken directly from the latter study and we refer the reader to that paper for details on the construction of the templates and derivation of the photometric data points. 

\begin{figure}
\center
 \includegraphics[width=\columnwidth]{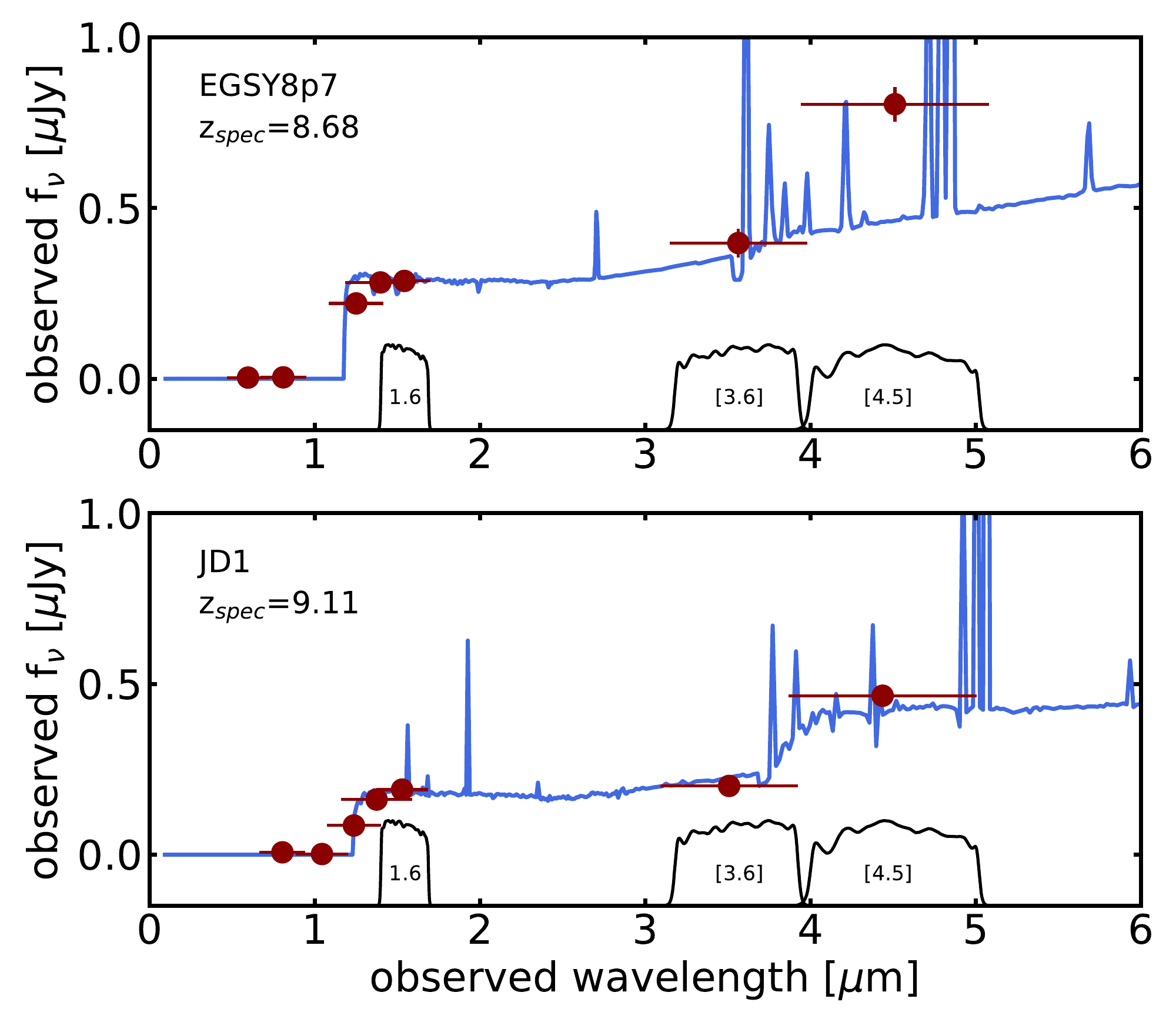}
 \caption{The two synthetic spectra (blue lines) and the associated observed photometry (dark red points with error bars; \citealt{rb16,zheng17}) used in this study to demonstrate the similarity of the redshift-dependent \textit{Spitzer}/IRAC [3.6]$-$[4.5] evolution. The spectra are of EGSY8p7 \citep{rb16,zitrin15}, a supposed extreme [\OIII]+H$\beta$ line emitter at $z=$8.68 (top), and JD1 \citep{hashimoto18}, a Balmer break galaxy at $z=$9.11 (bottom). The JD1 spectrum shown here is uncorrected for magnification. Shown at the bottom of each panel are the $HST/H_{160}$ and $Spitzer$/IRAC 3.6 $\mu$m and 4.5 $\mu$m response filters, for reference.}
 \label{fig:specs}
\end{figure}

Similarly, for the case of a mature stellar population with a prominent Balmer break, we select the SED fit to MACS1149-JD1 at $z$=9.11 from \citet{hashimoto18}. This fit represents a composite of a mature $\sim$200-300 Myrs population augmented with a younger component invoked to match the intensity of [\OIII] emission at 88 $\mu$m discovered with ALMA; full details can be found in \citet{hashimoto18}.  Since this spectral fit includes nebular line emission which, while not contributing significantly to the IRAC bands at $z>$9 will do so at lower redshift, we also explore the effect of suppressing all optical emission lines from the fitted spectrum of JD1. The observed (rest-frame) EWs of the combined [\OIII] and H$\beta$ lines are EW$_{\text{obs}}$([\OIII]+H$\beta$)$\approx$6690 \AA\ (EW$_{\text{rest}}$([\OIII]+H$\beta$)$\approx$770 \AA) and EW$_{\text{obs}}$([\OIII]+H$\beta$)$\approx$3408 \AA\ (EW$_{\text{rest}}$([\OIII]+H$\beta$)$\approx$375 \AA) for EGSY8p7 and JD1, respectively. The adopted spectral templates and associated photometry are shown in Figure \ref{fig:specs}, and for all subsequent analysis, we normalise both spectra by their flux at 0.325 $\mu$m (rest-frame), where the spectra are free from emission or absorption features, in order to ensure their 3.6 $\mu$m and 4.5 $\mu$m photometry can be directly compared.

\begin{figure}
\center
 \includegraphics[width=\columnwidth]{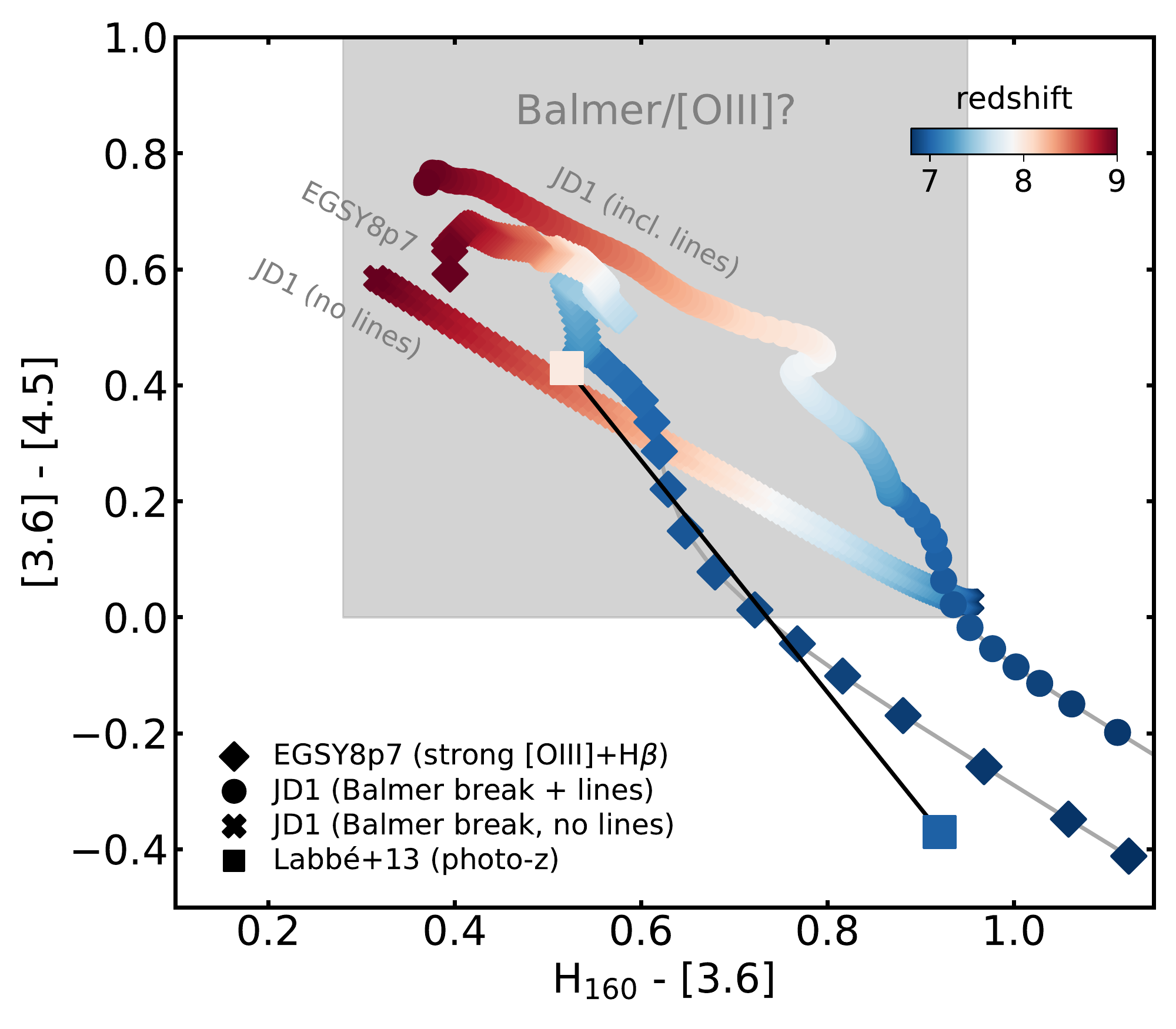}
 \caption{The [3.6]$-$[4.5] vs H$_{160}-$[3.6] colours as a function of redshift from 6.8$\leqslant z\leqslant$9 for the spectral templates of spectroscopically-confirmed galaxies EGSY8p7 (diamonds; assumed to be an extreme [\OIII]+H$\beta$ emitter) and JD1 (circles; a Balmer break galaxy) and the average SED results of \citet{labbe13} (squares). Each are connected by gray lines or a black vector for clarity. Since the template of JD1 includes some optical nebular emission, we also show the results for JD1 with masked emission lines (crosses) to highlight the contribution from the stellar continuum only. The gray shaded region indicates the colour space over which the different spectra produce similar colours for the selected redshift interval. A clear difference is evident between spectroscopic and photometric results, for which blue \textit{Spitzer}/IRAC colours at $z_{\text{phot}}\sim$7 are likely driven by strong nebular line emitters at $z\sim$6.6-6.8.}
 \label{fig:labbe}
\end{figure}

We are now in a position to explore how these template spectra affect the IRAC colours over the redshift range 6.8$<z<$9. In Figure \ref{fig:labbe} we present the redshift evolution of the [3.6]$-$[4.5] and $H_{160}-$[3.6] colours in steps of $\Delta\,z=0.02$ for each of our fiducial spectra, following a similar approach by \citet{labbe13}. At each redshift interval, the colours are measured directly from the redshifted spectrum using the relevant filter response curves. The simulation shows that the Balmer break in JD1 can mimic the effect of intense line emission to within $\pm$0.1 mag in the [3.6]$-$[4.5] colour, particularly at redshifts $z\gtrsim$7.5, and even produce redder colours at $z\gtrsim$8.5. Since the JD1 template includes a contribution from nebular emission lines (e.g., [\OII], H$\beta$ and [\OIII]), including some from H$\beta$ and [\OIII] $\lambdaup$4959 \AA\ at $z$=9.11 in the 4.5 $\mu$m band, we explored suppressing all optical line emission in this template but find consistently red IRAC colours with virtually identical colour-colour evolution (particularly at $z\gtrsim$8) and little difference in normalisation: the masked spectrum consistently produces red IRAC [3.6]$-$[4.5] colours $\lesssim0.2$ mag lower than its unmasked counterpart) on the simulated colours (see discussion below and Figure \ref{fig:labbe}).

The similarity between the redshift-dependent trends of a Balmer break and intense line emission over the chosen redshift interval 7$<z<$9 may seem surprising given earlier conclusions of a similar exercise undertaken by \citet{labbe13}. Those authors explored the effect by analysing average SEDs (containing nebular emission lines) of $z_{\text{phot}}\approx$7 and $z_{\text{phot}}\approx$8 galaxies selected as Lyman-break ``dropouts'' in the HUDF09+ERS and CANDELS-GOODS South fields. They were able to reproduce the average [3.6]$-$[4.5] and $H_{160}-$[3.6] colour differences only through the inclusion of high EW nebular emission lines (EW([\OIII]+H$\beta$)$\sim$560-670 \AA), rejecting an increased dust content or stellar age as major contributors to the colour evolution. We show their average $z\sim7$ and $z\sim8$ colours together with our results in Figure \ref{fig:labbe}.

However, while samples of $z\sim$8 galaxies selected as $Y$-dropouts over the aforementioned fields have average photometric redshifts of $z\sim$7.9 \citep{bouwens11,oesch12,bouwens15}, $z\sim7$ galaxies selected as $z_{850}$ dropouts have average redshifts of $z\sim$6.8 \citep{bouwens15}. As shown by \citet{smit15} and \citet{rb16}, the \textit{Spitzer}/IRAC [3.6]$-$[4.5] colour changes dramatically from $z\approx$7$\,\to\,$6.8 due to the strong [\OIII]+H$\beta$ lines transiting between filters, resulting in a change from especially red to especially blue colours. Thus, the blue colours observed in the \citet{labbe13} $z\sim7$ stack are conceivably due to the presence of $z\sim$6.8 galaxies which boost the $z=$7$\,\to\,$8 IRAC colour difference (as is clearly illustrated by our spectroscopic results where blue colours are apparent only at $z<$7 and red colours at $z\geqslant$7), thereby requiring more intense contributions to the colour evolution. The actual colour difference is in fact much smaller ($\sim$0.3-0.4 mag as opposed to $\sim$0.8 mag, see Figure \ref{fig:labbe}) and consistent with the trends observed for both strong nebular emission and Balmer break SEDs.

At this point we caution the reader that this exercise is not motivated to claim that the IRAC excess seen in many sources at $7<z<9$ cannot be due to intense [\OIII]+H$\beta$ emission. Rather, we wish to point out that the existence of a Balmer break for JD1 at $z>9$ may imply {\it some contribution of starlight to the IRAC excess} seen in sources at $7<z<9$ and to explore whether such starlight is prominent in existing spectroscopically-confirmed galaxies at $7<z<9$.  Of course, these results will depend on the choice of spectral template. In the case of JD1 we selected the only case known to date of a galaxy with an IRAC excess which cannot be explained solely via intense line emission \citep{hashimoto18}. For the line emitter template, the results will differ slightly depending on which object is chosen from the samples available in the literature. Given this uncertainty plus recent discussions on the possible exceptional case of JD1 \citep{bingelli2019}, it is helpful to examine the redshift-dependent trends produced by stellar population synthesis models as well as to expand the discussion to compare all these predictions with actual $z>7$ data in the literature. 

For the population synthesis models we use the \pegase\ suite \citep{fioc19} which includes self-consistent modelling of nebular line emission and dust evolution. Clearly such models have an abundance of free parameters but, for the present exercise, our main goal is to demonstrate that \textit{relative contributions} of synthetic galaxy spectra selected at various ages from a simple star formation history can also reproduce the trends we see using our observed spectral templates. For the current experiment, our simulated galaxy adopts a \citet{chabrier03} IMF with a constant star formation history beginning at $z=15$ and ending at $z=2$. We select simulated spectra at various time intervals corresponding to galaxy ages from 1 Myr ($z\sim$15) to 600 Myr ($z\sim$6). Nebular emission and dust evolution are included in the modelling and, for the former, we accommodate the possibility of a multiplicative factor L$_{\text{neb}}$ for the absolute strength of the lines in order to allow for extreme emission. As with the EGSY8p7 and JD1 spectra, the \pegase\ spectra are normalised to their flux at 0.325 $\mu$m (rest-frame) prior to analysis.

\begin{figure*}
\center
 \includegraphics[width=1.15\columnwidth]{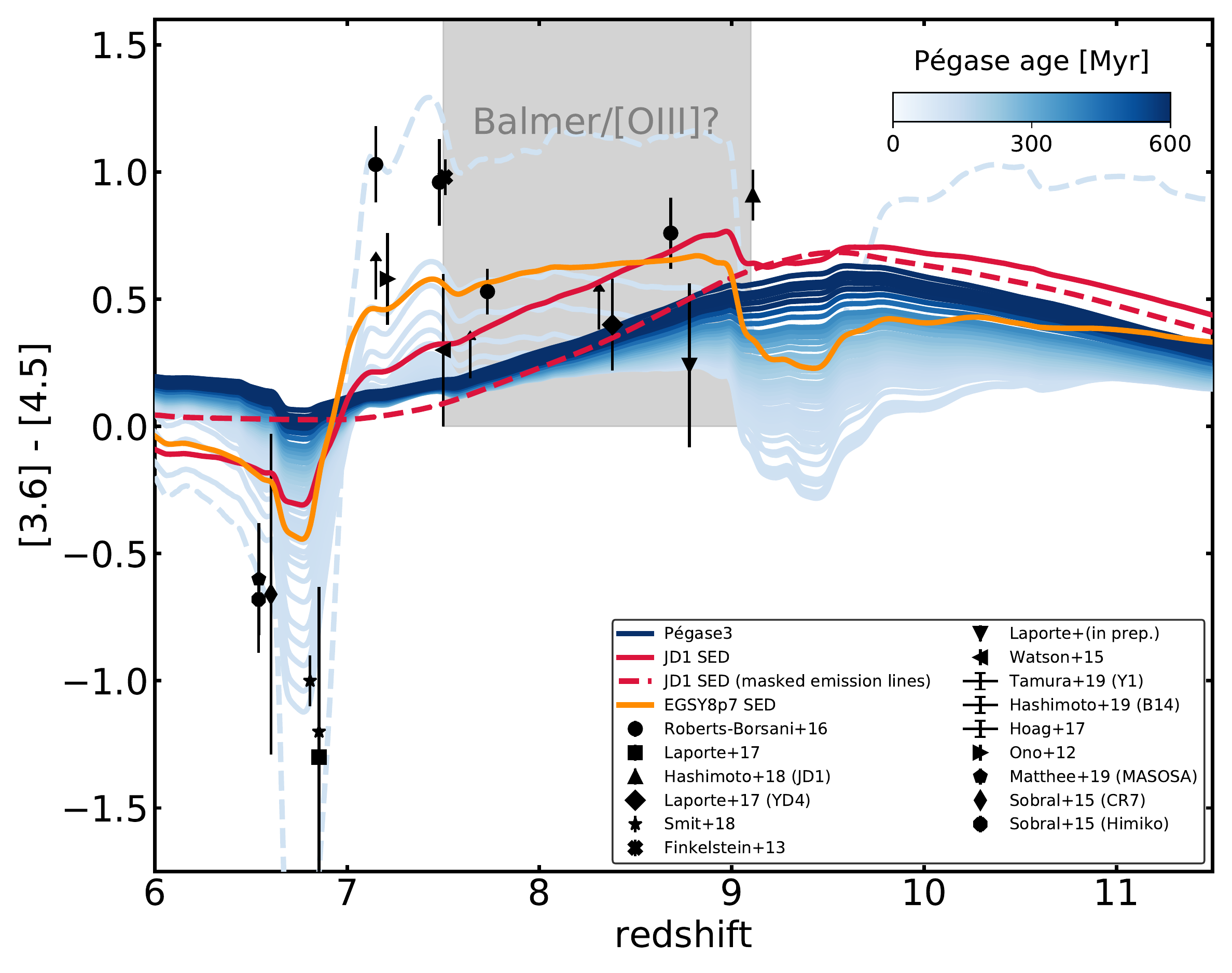}
 \caption{The \textit{Spitzer}/IRAC [3.6]$-$[4.5] colour evolution as a function of redshift. Black points represent spectroscopically-confirmed galaxies at $z>$7 with red IRAC colours (see Table \ref{tab:z7}) with additional data for $z\sim$6.6-6.8 sources with especially blue IRAC colours ([3.6]$-$[4.5]$<-$0.5). The evolution for the synthetic spectra of EGSY8p7 (orange), JD1 (red, solid), JD1 with masked emission lines (red, dashed) can be compared with synthesis models including normal (L$_{\text{neb}}$=1) line emission from \pegase\ (blue shades, solid lines) for a variety of galaxy ages, assuming a constant SFR from $z=$15 to $z=$2. The dashed blue line illustrates the colour evolution for the youngest \pegase\ spectrum and L$_{\text{neb}}$=5, which is sufficient to reproduce the most extreme colours of spectroscopically-confirmed galaxies. The light gray shaded region highlights an approximate redshift interval over which ambiguity exists as to the primary mechanism for red \textit{Spitzer}/IRAC colours.}
 \label{fig:colevol}
\end{figure*}

Figure \ref{fig:colevol} shows the IRAC [3.6]$-$[4.5] colour versus redshift trend for our fiducial galaxy templates as well as the \pegase\ models. For each of our SEDs (i.e., EGSY8p7, JD1 and each of the \pegase\ spectra corresponding to various time intervals of the simulated galaxy's evolution), the spectrum is redshifted across our range of interest and the colour measured through the relevant response filters, in order to assess the relative colour contributions from each spectrum's features. To illustrate the effects of adopting normal (L$_{\text{neb}}$=1) and extreme (L$_{\text{neb}}>>$1) nebular emission for, respectively, this portion of the analysis and that adopted later, we also overplot the youngest galaxy spectrum with L$_{\text{neb}}$=5, which is sufficient to match even those most extreme blue and red colours at $z\sim$6.6-6.8 and $z\sim$7-7.5, respectively. Additionally, the figure also shows the JD1 template adjusted to exclude the contribution from optical nebular lines. In order to compare with actual data, Table \ref{tab:z7} represents a compilation of 13 $z>$7 spectroscopically-confirmed galaxies drawn from the literature, each with available \textit{HST} and \textit{Spitzer}/IRAC photometry, and we plot their photometric data alongside the spectroscopic results in Figure \ref{fig:colevol}. For completeness we also add spectroscopically-confirmed sources at $z\simeq$6.8 sources with especially blue ([3.6]$-$[4.5]$<-$0.5 mag) colours \citet{sobral15,laporte17b,smit18,matthee19} which demonstrate the influence of [\OIII]+H$\beta$ emission in the 3.6 $\mu$m band at lower redshift.

Focusing initially on the comparison between \pegase\ and our chosen galaxy templates, we can see very similar trends, albeit with some difference in normalisation. Within the $7<z<9$ redshift range, P\'egase models corresponding to younger ages closely track the evolution of the EGSY8p7 template, whose red colours are dominated by strong nebular emission lines, whilst evolved stellar ages are required to explain the evolution of the JD1 templates (both with and without emission lines), where the red colour is primarily due to a Balmer break.

Considering next how the templates and synthesis models match the IRAC colours of 13 spectroscopically-confirmed $z>7$ galaxies, we can see that line emission in both \pegase\ and the EGSY8p7 (line emitting) template are required to explain the strong dip in [3.6]$-$[4.5] colour at $z\simeq$6.6-6.8 as indicated by \citet{smit18}; as expected the JD1 template with masked emission lines has no dip. However, at higher redshift, where the [\OIII]+H$\beta$ lines enter the 4.5 $\mu$m band, the red colour is initially more easily reproduced by cases with strong line emission. For SEDs with a flat continuum, as assumed by \citet{rb16}, such a colour remains relatively constant until the lines leave the band at $z=9$. However, in the case of a moderate to strong Balmer break and reduced (but not absent) line emission, the IRAC excess \textit{increases} from $z\gtrsim$7.5 onwards as the Balmer break moves redward, thereby removing flux from the 3.6 $\mu$m band whilst simultaneously providing a relatively constant amount of flux in the 4.5 $\mu$m band. In the case where the rest-frame optical continuum is not flat, this impacts the \textit{slope} of the [3.6]$-$[4.5] colour evolution. The impact of the Balmer break is particularly evident when comparing the evolution of the masked JD1 spectrum to the older-aged \pegase\ synthesis models which, despite including (normal) emission lines display virtually identical [3.6]$-$[4.5] evolution.

Examining the actual data, one can reasonably securely conclude that the large IRAC excesses at $7\lesssim z \lesssim7.5$ (e.g. \citealt{ono12,finkelstein13,rb16,hashimoto19}) are difficult to reproduce without an extreme [\OIII]+H$\beta$ contribution, as is clearly the case for IRAC colours for the sources at $z\simeq$6.8 studied by e.g., \citet{sobral15}, \citet{smit18}, \citet{laporte17b} and \citet{matthee19}. However, for the sources at $7.5\lesssim z\leqslant9$ (e.g., \citealt{watson15,hoag17,laporte17,tamura19} plus GN-z10-3 and EGSY8p7), the paucity of spectroscopic data makes it premature to conclude that the IRAC excess in galaxies beyond $z\simeq$7 arises entirely from line emission.

\begin{table*}
    \centering
    \begin{tabular}{lcccr}
       ID  &  $z_{\text{spec}}$ & H$_{160}$ & 3.6 $\mu$m$-$4.5 $\mu$m & Reference\\ \hline
       MACS1149\_JD1  & 9.11  &  25.70$\pm$0.01  & 0.91$\pm$0.10 & \cite{hashimoto18} \\
       GN-z10-3       & 8.78  &  26.74$\pm$0.12  & 0.24$\pm$0.32 & Laporte et al. (in prep.) \\
       EGS8p7         & 8.68  &  25.26$\pm$0.09  & 0.76$\pm$0.14 & \cite{zitrin15} \\
       A2744\_YD4     & 8.38  &  26.42$\pm$0.04  & 0.40$\pm$0.18  & \cite{laporte17} \\
       MACS0416\_Y1   & 8.31  &  26.04$\pm$0.05  & $>$0.38       & \cite{tamura19} \\
       EGS-zs8-1      & 7.73  &  25.03$\pm$0.05  & 0.53$\pm$0.09 & \cite{oesch15} \\
       MACS1423-z7p64  & 7.64  &  25.03$\pm$0.10  & $>$0.19 & \cite{hoag17} \\
       z8\_GND\_5296  & 7.51  &  25.55$\pm$0.07  & 0.98$\pm$0.07 & \cite{finkelstein13} \\
       A1689-zD1      & 7.50  &  24.70$\pm$0.10  & 0.30$\pm$0.30 & \cite{watson15} \\
       EGS-zs8-2      & 7.48  &  25.12$\pm$0.05  & 0.96$\pm$0.17 & \cite{rb16,stark17} \\
       GN-108036      & 7.21  &  25.17$\pm$0.07  & 0.58$\pm$0.18 & \cite{ono12} \\
       COSY           & 7.15  &  25.06$\pm$0.06  & 1.03$\pm$0.06 & \cite{stark17} \\
       B14-65666      & 7.15  &  24.60$^{+0.3}_{-0.2}$ & $>0.5$  & \cite{hashimoto19} \\
      \hline 
    \end{tabular}
    \caption{A list of spectroscopically-confirmed galaxies at $z>$7 with red IRAC colours.}
    \label{tab:z7}
\end{table*}

\begin{table*}
    \centering
    \begin{tabular}{lccccccc}
      ID  & $z_{\text{spec}}$ & $\mu$ & $\lambdaup_{\text{[\OIII]}}$ & [\OIII] integrated flux & [\OIII] FWHM & $\lambdaup_{\text{cont}}$ & continuum flux \\
        &  &  & [$\mu$m] & [Jy\,km/s] & [km/s] & [$\mu$m] & $\mu$Jy\\ \hline
      MACS1149\_JD1  & 9.11 & 10 & 893.86 & 0.229$\pm$0.050 & 154$\pm$39 & 917.63 & $<$35.4\\
      A2744\_YD4     & 8.38 & 2 & 829.55 & 0.030$\pm$0.008 & 49.8$\pm$4.2 & 842.70 & 99.0$\pm$23.0\\
      MACS0416\_Y1   & 8.31 & 1.43 & 823.32 & 0.660$\pm$0.160 & 141$\pm$21 & 850.57 & 137.0$\pm$26.0\\
      B14-65666      & 7.15 & -- & 720.78 & 1.500$\pm$0.180 & 429$\pm$37 & 733.68 & 470.0$\pm$128.0\\
      \hline 
    \end{tabular}
    \caption{A list of $z>7$ galaxies from Table \ref{tab:z7} with [\OIII] $\lambdaup$88 $\mu$m detections and dust continuum constraints (detections and non-detections) from ALMA Band 7 observations. The columns represent, from left to right: the redshift of the galaxy, the assumed magnification factor, the observed wavelength of the [\OIII] $\lambdaup$88 $\mu$m detection, its integrated flux, and the measured FWHM. The last two columns are the central observed wavelength of the dust continuum observations and the associated flux. All error bars and upper limits quoted here are 2$\sigma$.}
    \label{tab:z7oiii}
\end{table*}

\section{Distinguishing Between a Balmer break and Intense Line Emission}
\label{sec:distinguish}

\subsection{ALMA data and the origin of the [\OIII] line ratio}
\label{subsec:almaoiii}
In Section \ref{sec:iracmodelling} we have shown that red \textit{Spitzer}/IRAC [3.6]$-$[4.5] colours for galaxies lying between $7<z<9$ could arise from contributions of both intense nebular line emission and starlight. We now consider whether it is possible to break this degeneracy prior to the use of spectroscopy with JWST. ALMA observations with Band 7 targeting the [\OIII] $\lambdaup$88 $\mu$m line and dust continuum may provide a potential way forward. [\OIII] $\lambdaup$5007 \AA\ and [\OIII] $\lambdaup$88 $\mu$m emission originate from the same star-forming regions and species, and thus constraints from the [\OIII] $\lambdaup$88 $\mu$m line should provide valuable limits to the strength of the [\OIII] $\lambdaup$5007 \AA\ emission and therefore \textit{Spitzer}/IRAC [4.5] flux contributions. Furthermore, while the precise origin and nature of dust at high-$z$ remains an open debate, constraints from continuum emission can place additional (indirect) constraints on the rate of supernovae explosions and and thus the underlying stellar populations. Using detailed SED modelling with synthetic spectra from \pegase, we now investigate whether ALMA observations can place constraints on the contribution from young stars via [\OIII] $\lambdaup$88 $\mu$m emission and mature stellar populations from the presence of a dust continuum.

Currently, four of the 13 $z>$7 spectroscopically-confirmed galaxies with \textit{Spitzer}/IRAC excesses listed in Table \ref{tab:z7} have the appropriate ALMA data: JD1, A2744\_YD4, MACS0416\_Y1 and B14-65666 (henceforth YD4, Y1 and B14 for convenience), see Table \ref{tab:z7oiii}. To determine accurate SEDs, for each of the aforementioned galaxies we use the relevant references in Table \ref{tab:z7} (and references therein) to compile \textit{HST} ACS+WFC3/IR, ($B_{\text{435}}$, $V_{\text{606}}$, $I_{\text{814}}$, $Y_{\text{105}}$, $J_{\text{125}}$, $JH_{\text{140}}$ and $H_{\text{160}}$ bands), VLT/HAWK-I \textit{K$_{s}$}, as well as \textit{Spitzer}/IRAC [3.6] and [4.5] photometry. For B14 we use near-infrared $z$, $Y$, $J$, and $H$ photometry from VISTA, in addition to data in the VLT/HAWK-I \textit{K$_{s}$} and \textit{Spitzer}/IRAC [3.6] and [4.5] bands. All \textit{HST} upper limits and error bars represent 1$\sigma$ uncertainties, whilst those redward of these are 2$\sigma$.

To evaluate the relative contributions of nebular emission lines and starlight we create a repertoire of \pegase\ spectra with which to fit the above data for the four spectroscopically-confirmed $z>$7 galaxies. We generate mass-normalised galaxy spectra for a young component dominated by a recent burst of constant star formation with duration $\tau_{\text{young}}$=10 Myrs, and for a component with a less recent phase of constant star formation for a range of durations $\tau_{\text{old}}$=[10, 50, 100, 200, 300, 400, 500] Myrs, where a Balmer break is allowed to form. We then extract spectra at 1 Myr intervals for the young component, and 20 Myr intervals from ages of 1 Myr to the age of the Universe at the redshift of each galaxy for the older component. For simplicity, we assume emission lines arise from the young component only, since these come from star-forming regions and are not seen in mature stellar populations. These models are used, sometimes in combination (i.e. recent burst + earlier star formation), with a custom SED-fitting code in a Bayesian framework, to maximise the log-likelihood of the model given the data, including an analytical treatment of upper limits \citep{sawicki12}. The free parameters of the code are the mass of the galaxy system, M$_{\text{sys}}$\footnote{For a proper definition of how this translates in the \pegase\ formalism to stellar mass, see \citealt{fioc19}.} (one for each stellar component) and a multiplicative factor, L$_{\text{neb}}$, to scale the luminosity of the nebular emission lines, whose FWHM are fixed to that of the [\OIII] $\lambdaup$88 $\mu$m line presented in Table \ref{tab:z7oiii}. The adopted priors are log M$_{\text{sys}}$=[5,15] M$_{\odot}$ and L$_{\text{neb}}$=[0,50], allowing for both normal and extreme nebular emission contributions.

Throughout the subsequent analysis, we make the assumption that the [\OIII] $\lambdaup$5007 \AA\ line can be constrained from the flux of the [\OIII] $\lambdaup$88 $\mu$m line, since these originate from the same star-forming regions. However, the ratio of the two lines is highly dependent on the conditions of the surrounding gas (e.g., electron density and temperature, gas metallicity, the emission rate of ionising photons from the ionising star), given that the two species are characterised by different excitation energies and critical densities. We therefore provide a short description of \pegase's nebular emission treatment (for full details we refer the reader to \citealt{fioc19}) and consider the [\OIII] ratios probed by our young starburst component to ensure they are suitable for a meaningful analysis of the emission line contributions to the \textit{Spitzer}/IRAC channels.

\pegase\ determines the integrated luminosities of a large suite of nebular emission lines by using a pre-computed luminosity grid from \texttt{Cloudy} \citep{ferland17} and linking it to the stellar and ISM evolution of the galaxy model. The large emission grid was constructed as a function of ISM metallicity and emission rate of Lyman continuum photons (with ranges \{0-0.1\} and \{10$^{46}$-10$^{53}$\} s$^{-1}$ respectively) for spherically symmetric, radiation bound \HII\ regions filled with a constant gas density of $n_{\text{H}}$=10$^{2}$ cm$^{-3}$. Each of the \texttt{Cloudy} simulations were run out to a radial distance from the central ionising star where the free proton density drops below 10$^{-2}$ cm$^{-3}$. For each star cluster in a given Simple Stellar Population (SSP) with age $t$, \pegase\ computes the ionising output and gas phase metallicity prior to determining the integrated line luminosity through interpolatation of the \texttt{Cloudy} grid. The emission lines from \pegase\ are thus computed in a self-consistent manner and directly linked to the underlying stellar and ISM evolution. As such, no prior assumptions on either the electron density or temperature need be supplied.

\begin{figure}
\center
 \includegraphics[width=\columnwidth]{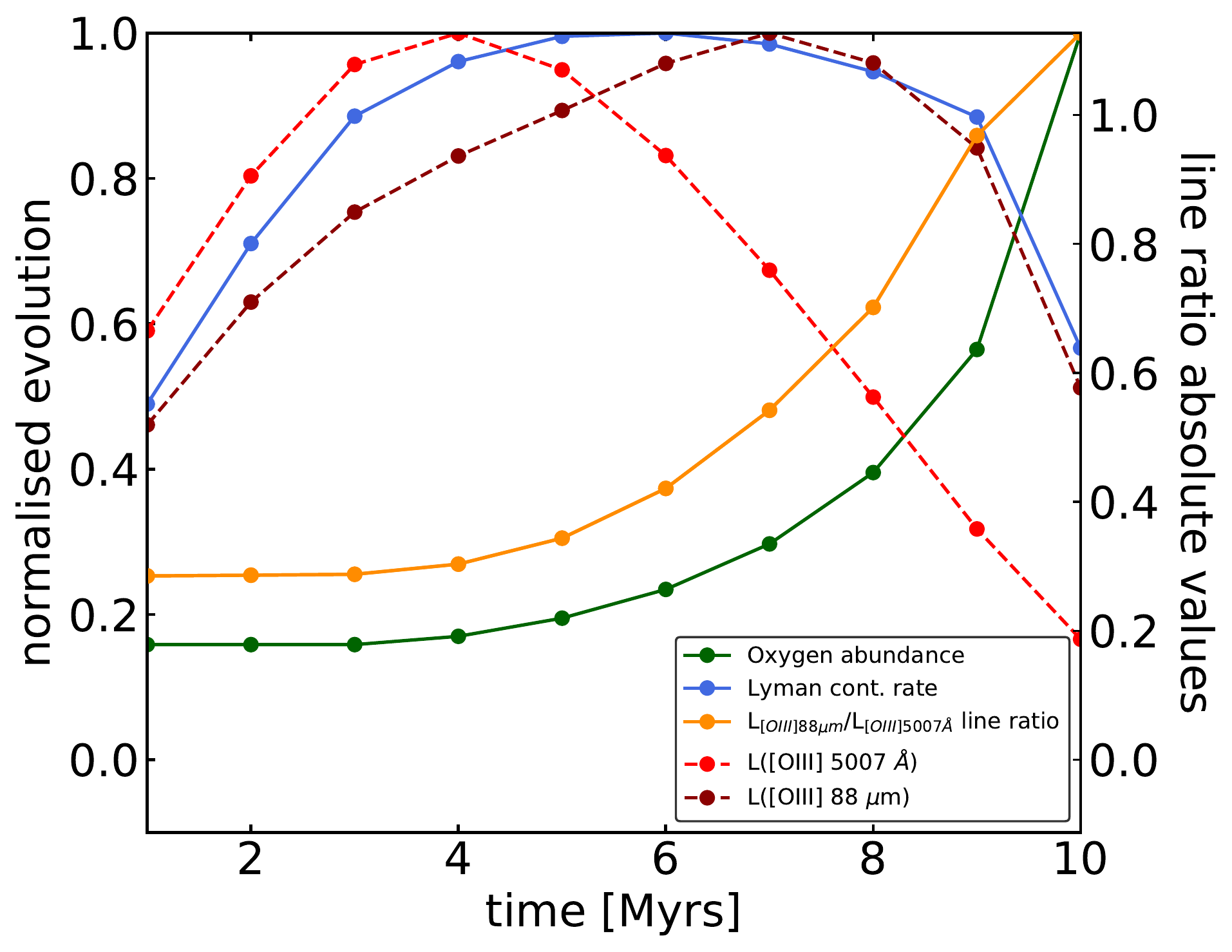}
 \caption{The evolution of the [\OIII] $\lambdaup$88 $\mu$m / [\OIII] $\lambdaup$5007 \AA\ line ratio as a function of time for our young burst model, with associated galaxy properties (Oxygen abundance, Lyman continuum emission rate and the luminosities of both lines). Each quantity is normalised by its maximum value and a  clear trend is seen between the line ratio and the galaxy metallicity, due to cooling effects and the large difference in excitation potential between the two lines. Our low ratios (i) are tied to the underlying stellar and gas evolution of the starburst model and (ii) probe a suitable range of ratios that favour intense [\OIII] $\lambdaup$5007 \AA\ nebular emission capable of producing up to $\sim$1.4 mag \textit{Spitzer}/IRAC [3.6]$-$[4.5] colours.}
 \label{fig:ratio}
\end{figure}

With the above in mind, we plot in Figure \ref{fig:ratio} the normalised line ratio as a function of time for the young starburst model, in addition to several relevant galaxy quantities. Our [\OIII] $\lambdaup$88 $\mu$m / [\OIII] $\lambdaup$5007 \AA\ values span a reasonably large range of \{0.28$-$1.13\}. Using rest-frame optical line ratios from our spectra as electron temperature ($T_{e}$) and density ($n_{e}$) diagnostics \citep{proxauf14}, we find our nebular emission also probe relatively low values of $T_{e}\sim$8,500-11,000 $\SI{}{\kelvin}$ and $n_{e}\lesssim$10$^{2}$ cm$^{-3}$, respectively. Given such values remain well below the critical densities of the two [\OIII] lines (510 cm$^{-3}$ and 6.8$\times$10$^{5}$ cm$^{-3}$ for the 88 $\mu$m and 5007 \AA\ line, respectively; \citealt{osterbrock06}), collisional de-excitation is unlikely to play a major role in regulating the luminosities of the lines. Instead, the large difference between the excitation potentials of the two lines (0.014 eV and 2.476 eV for the 88 $\mu$m and 5007 \AA\ line, respectively) suggests temperature fluctuations primarily due to the gas-phase metallicity acts as the primary regulator of the ratio, as is clearly illustrated in the figure. At very low metallicities, the increase in Lyman continuum photon production rate coincides with a rise in temperature and thus the number of collisions with energetic electrons and population of the [\OIII] $\lambdaup$5007 \AA\ upper energy level. Given the low excitation potential of the [\OIII] $\lambdaup$88 $\mu$m line however, its luminosity remains largely unaffected and thus does not vary significantly, maintaining a low line ratio. Once the production of metals becomes significant, however, cooling effects begin to take place and the luminosity of the [\OIII] $\lambdaup$5007 \AA\ line drops dramatically, thereby rapidly increasing the line ratio towards unity.

Prior to the launch of JWST, observations of the [\OIII] $\lambdaup$5007 \AA\ line at $z>$6 are not possible, while the number of low-$z$ analogues with matched observations of both lines is limited. Since metallicity plays a key role in regulating the evolution of the line ratio, however, we compare the Oxygen abundances for each of our young starburst spectra (and therefore line ratios) to those of the local, low-metallicity dwarf galaxies in the Dwarf Galaxy Survey (DGS; \citealt{madden13}). To be consistent with the metallicity derivations of the DGS, we use the ``R23'' ([\OII] $\lambdaup\lambdaup$3727,3729 $+$ [\OIII] $\lambdaup\lambdaup$4959,5007)/H$\beta$ parameterisations by \citet{pilyugin05} with a log([\NII] $\lambdaup$6584/[\OII] $\lambdaup\lambdaup$3727,3729)$\sim-$1.2 limit in order to break the R23 degeneracy \citep{kewley08}. Our derived values range from 12$+$log (O/H)$\approx$8.40$-$8.67 (0.50-0.93 $Z_{\odot}$), characteristic of sub-solar abundances and consistent with the high end of the DGS metallicities.

Finally, we find our range of line ratios more than capable of producing extreme rest-frame optical emission, with $z>$7 \textit{Spitzer}/IRAC colours up to [3.6]$-$[4.5]$\sim$1.4 mag. This is consistent with all of the red colours of $z>$7 spectroscopically-confirmed galaxies in Figure \ref{fig:colevol} and much larger than those reported for the four galaxies selected above. Should our ratios have probed larger values only, characteristic of other areas in a simple $T_{e}-n_{e}$ grid, subsequent analyses would likely be biased against strong nebular emission. Thus, given the consistency of our line ratios with expected trends and data for low-metallicity systems, as well as the large \textit{Spitzer}/IRAC excess allowed by our models, we deem the nebular emission generated here well-suited for the purposes of the present paper.

\begin{figure*}
\center
 \includegraphics[width=1.75\columnwidth]{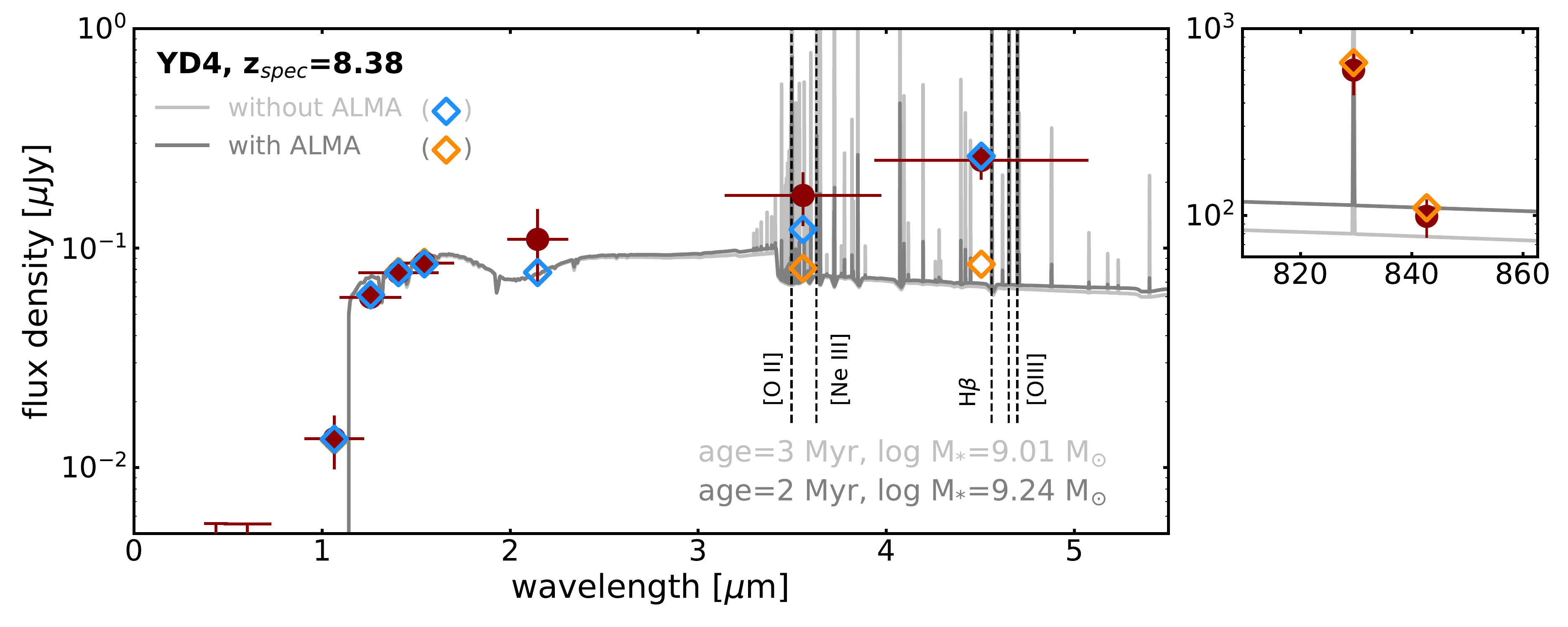}
 \caption{SED fits to the spectroscopically-confirmed galaxy, YD4, with \textit{HST}, VLT/HAWK-I \textit{K$_{s}$} and \textit{Spitzer}/IRAC photometry (red points and error bars) and single models of dusty 1-10 Myr \pegase\ spectra. The suite of young \pegase\ spectra are fit once without the ALMA Band 7 spectroscopic observations (light gray line and blue points) and once with them (dark gray line and orange points), both with the strength of nebular emission lines as a free parameter in addition to their stellar masses. The IRAC excess is well fit by contributions from nebular emission lines without inclusion of the ALMA data. However when such constraints are included, the nebular emission is suppressed and cannot account for the excess flux in the \textit{Spitzer}/IRAC 4.5 $\mu$m band. Thus, a secondary component arising from starlight is necessary to match the data.}
 \label{fig:validation}
\end{figure*}

To illustrate how the ALMA observations may differentiate between intense nebular emission and starlight in explaining the IRAC colours, we consider the case of YD4 since, for this source, all photometric points redward of the Lyman break and the ALMA spectroscopic constraints are robustly measured. First we fit the data with a single-component young model (permitting a dust contribution), once with \textit{HST}+VLT/HAWK-I+\textit{Spitzer}/IRAC data only and then again incorporating the ALMA Band 7 constraints. The best-fit SEDs are shown in Figure \ref{fig:validation}. The continuum fits to the \textit{HST} and HAWK-I photometry are satisfactory and comparable in each case. However, there is a major difference in the predicted \textit{Spitzer}/IRAC photometry. Ignoring the ALMA constraints, the IRAC excess demands the presence of strong nebular emission lines in which case  the [\OIII] $\lambdaup$88 $\mu$m line is considerably overpredicted (by a factor $\simeq$10, not shown) and the continuum dust emission is similarly poorly matched. Additionally, the presence of nebular emission lines - namely the [\OII] doublet at a rest-frame of $\sim$3730 \AA\ - adds non-negligible boosting to the [3.6] band. Including the ALMA constraints, the nebular emission in the IRAC bands is modest, indicating the need for an additional component to fit these data, for example a Balmer break originating from star formation at earlier times.

\subsection{SED-fitting \& Results}

We thus now proceed to fit all of the available data for each of the four galaxies with a two-component model comprising a contribution from young stars with intense nebular emission and an older stellar component. The two-component models are derived from all unique combinations of young and older spectra requiring only that age$_{\text{young}}<\text{age}_{\text{galaxy}}-\tau_{\text{old}}$ so that the recent burst of star formation from the young component occurs only when star formation in the older component has completed. In these two-component models, for simplicity, we assume the dust contributions arise only from the older component and specifically only if there is a Band 7 continuum detection. By comparing these two-component fits to those for a single-component (with dust included following the guidelines indicated above), we can determine, as suggested in the case of YD4, (i) whether the two-component fits are significantly better than the single young component ones and (ii) whether the first indicate the presence of an older, more mature stellar population.

\begin{figure*}
\center
 \includegraphics[width=1.25\columnwidth]{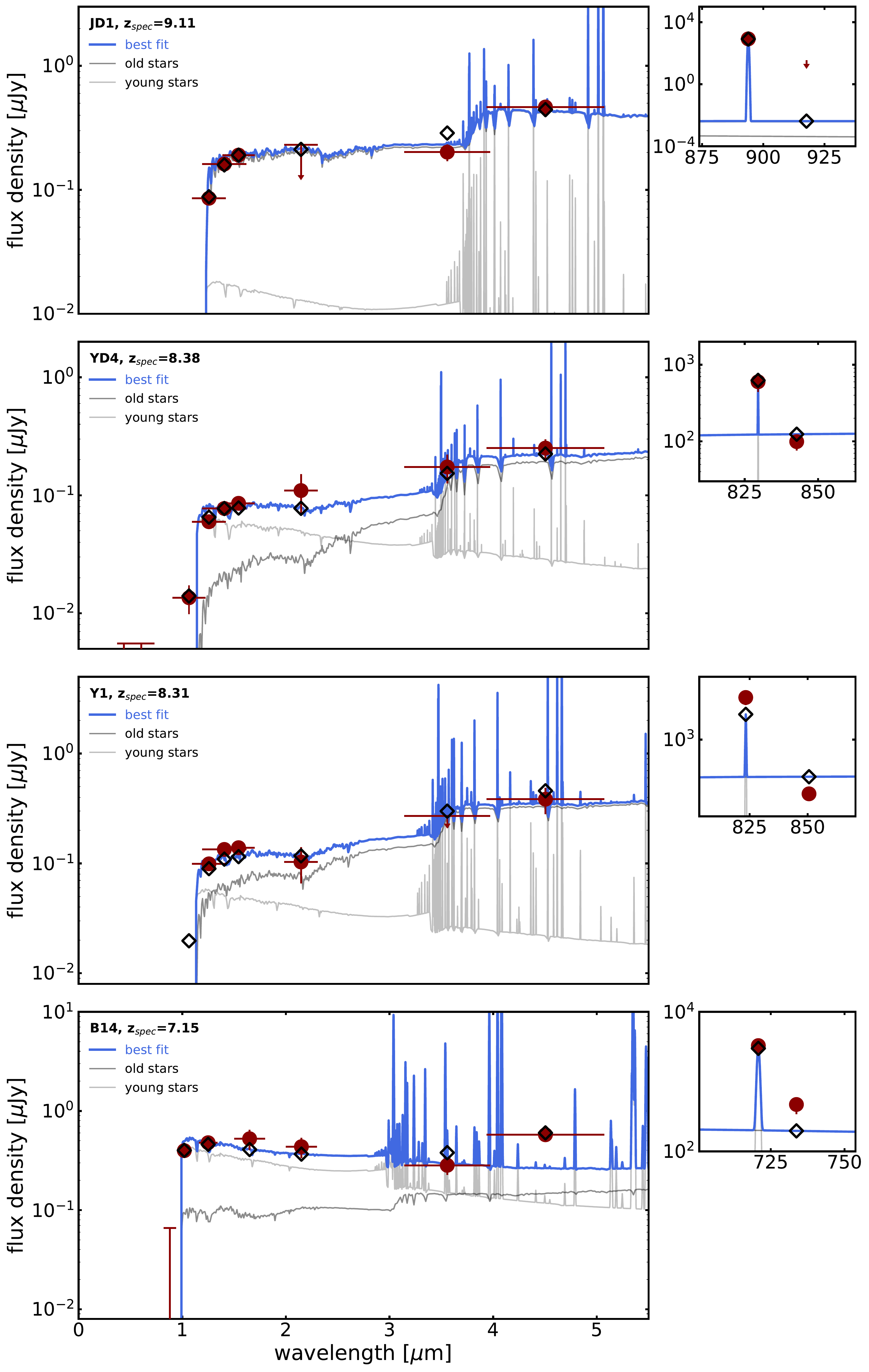}
 \caption{The best fit two-component SED models (blue lines and black points) to observed data (dark red circles and error bars) from \textit{HST}/VISTA + VLT/HAWK-I \textit{K$_{s}$} + \textit{Spitzer}/IRAC photometry (left) and ALMA spectroscopy (upper right inset) for galaxies at $z>7$ with red \textit{Spitzer}/IRAC colours (JD1, YD4, Y1 and B14). The total fit consists of contributions from both old (darker gray) and young (lighter gray) stars.}
 \label{fig:sedresults}
\end{figure*}

The results of our best fit two-component models are presented in Figure \ref{fig:sedresults}, where we find generally good agreement with the observed photometric data sets and ALMA constraints. For two of the four galaxies (JD1 and YD4), the best-fit model correctly predicts the presence or upper limit of dust mass based on the ALMA continuum detections, whilst simultaneously reproducing the \textit{HST} photometry, \textit{Spitzer}/IRAC excess, and [\OIII] $\lambdaup$88 $\mu$m flux, within the error bars. For Y1 and B14, on the other hand, the \textit{HST} photometry and \textit{Spitzer}/IRAC excess are well reproduced, but the models are unable to simultaneously match both the [\OIII] $\lambdaup$88 $\mu$m flux and dust continuum (in the case of B14, the [\OIII] $\lambdaup$88 $\mu$m flux is matched but the dust continuum is underpredicted, and in the case of Y1 the [\OIII] $\lambdaup$88 $\mu$m flux is underpredicted and dust continuum overpredicted). In comparing these fits to those assuming a single-component only, although the \textit{HST} photometry can be reasonably well-reproduced for JD1 and YD4, the single-component fits fail to simultaneously reproduce both the ALMA constraints and the \textit{Spitzer}/IRAC excess. In the cases of Y1 and B14, the ALMA constraints are better matched by the one-component fit, however the \textit{Spitzer}/IRAC excess is only partially matched.

In our two-component fits, a sizeable contribution to the IRAC flux arises from a more mature stellar component. For JD1, YD4 and Y1, the contribution to the IRAC fluxes from the recent burst of star formation is only $\sim$10-30\% and the older component (characterised by Balmer ratios of $\sim$2) dominates the flux at $\gtrsim$70\%. This is due primarily to the relative weakness of the [\OIII]+H$\beta$ lines, for which we measure an equivalent width, EW([\OIII]+H$\beta$)$\approx$25-106 \AA\ (compared to EW([\OIII]+H$\beta$)$\approx$450-7700 \AA\ for a single-component). The exception to this trend is B14, whose IRAC fluxes remain dominated by the younger component by $\sim$60-75\% (EW([\OIII]+H$\beta$)$\approx$200 \AA\ for the two-component model and EW([\OIII]+H$\beta$)$\approx$3460 \AA\ for the one-component model). The total stellar mass (corrected for lensing for those lensed sources) for all of these fits range from log M$_{*}$=8.88-10.19 M$_{\odot}$, with virtually all of the total stellar mass also coming from the earlier period of star formation and the most recent burst contributing primarily through the presence of weak-to-moderate nebular emission.

Furthermore, we also note a difference in the ages of the galaxies, as determined by the onset of the earlier burst in the two-component model or the age of the single-component model. We find an increase in age for each of the four galaxies when multiple components are used, with ages of 80, 260, 140 and 20 Myrs characterising the two-component fits of JD1, YD4, Y1, and B14, and such ages decreasing down to 1, 2, 1 and 3 Myrs assuming on a single burst of recent star formation. In the case of Y1, the age estimate from the two-component fit is in fact a lower limit, since we only have an upper limit for the \textit{Spitzer}/IRAC 3.6 $\mu$m photometry. We note here that our preferred $\tau_{\text{old}}$=10 Myrs value and the galaxy age estimate for JD1 are somewhat lower than those estimated by \citet{hashimoto18}, whose best fit comprises an older stellar population with an episode of star formation lasting $\tau_{\text{old}}$=100 Myrs and a galaxy age of 290$^{+190}_{-120}$ Myrs. However, as in their analysis, we find considerable statistical similarity with the best two-component fit assuming $\tau_{\text{old}}$=100 Myrs, in which case our age estimate increases to 140 Myrs, closer to the lower limit of their assumed value.

Finally, to enable a quantitative comparison of the one- and two-component fits we examine the log-likelihoods and find, with the exception of Y1, that the two-component fits of each of the galaxies is considered a better fit. However, given the obvious danger of concluding better fits with an additional component with further free parameters, we compare the goodness-of-fit via a comparison of their Bayesian Information Criteria (BIC), which uses the log-likelihoods whilst penalising for additional free parameters. With this consideration, the additional free parameters in the fit to the B14 data are sufficiently penalised to justify only the one-component fit. Whilst such comparisons no doubt gloss over the complexities of defining the birth of such galaxies, they serve as an important illustration of the potential consequences from overlooking the consideration of multiple episodes of star formation. We provide a summary of the above comparison and the main properties of our fits in Table \ref{tab:sedres}.

\begin{table*}
    \centering
    \begin{tabular}{lcccccccc}
    \hline
     & \multicolumn{2}{c}{JD1} & \multicolumn{2}{c}{YD4} & \multicolumn{2}{c}{Y1} & \multicolumn{2}{c}{B14}  \\ \hline
       & 1-comp & 2-comp & 1-comp & 2-comp & 1-comp & 2-comp & 1-comp & 2-comp  \\ \hline
      $\tau_{\text{old}}$ [Myrs] & -- & 10 & -- & 200 & -- & 100 & -- & 10 \\
      galaxy age [Myrs] & 1 & 80 & 2 & 260 & 1 & 140 & 3 & 20 \\
      M$_{*,\text{young}}$/M$_{*,\text{old}}$ & -- & 2$\times$10$^{-3}$ & -- & 3.2$\times$10$^{-3}$ & -- & 1.9$\times$10$^{-3}$ & -- & 8$\times$10$^{-2}$ \\
      log M$_{*,\text{total}}$ [M$_{\odot}$] & 8.23$\pm$0.01 & 9.88$\pm$0.01 & 8.71$\pm$0.02 & 10.26$\pm$0.06 & 8.93$\pm$0.01 & 10.35$\pm$0.03 & 9.25$\pm$0.04 & 9.49$\pm$0.99 \\
      EW([\OIII]+H$\beta$) [\AA] & 7668$\pm$793 & 106$\pm$43 & 448$\pm$133 & 25$\pm$9 & 7307$\pm$1211 & 51$\pm$11 & 3462$\pm$720 & 198$\pm$260 \\
      D$_{n}$4000 & 0.55$\pm$0.01 & 1.94$\pm$0.02 & 0.77$\pm$0.00 & 2.09$\pm$0.07 & 0.81$\pm$0.01 & 1.99$\pm$0.02 & 0.82$\pm$0.01 & 0.97$\pm$0.35 \\
      \hline
      3.6 $\mu$m flux contribution & & (9.8, 90.2) \% &  & (25.4, 74.6) \% &  & (23.9, 76.1) \% &  & (62.4, 37.6) \% \\
      4.5 $\mu$m flux contribution & & (8.2, 91.8) \% &  & (18.4, 81.6) \% &  & (30.8, 69.2) \% &  & (75.5, 24.5) \% \\
      \hline
      log-likelihood & -188.81 & -12.22 & -9.44 & -3.02 & -62.47 & -66.92 & -6.17 & -4.86 \\
      BIC & 403.21 & 62.84 & 44.02 & 43.76 & 150.33 & 171.94 & 38.10 & 48.37 \\
      \hline 
    \end{tabular}
    \caption{Summary of the main properties and parameters (uncorrected for any lensing of the objects) of the favoured one- and two-component SED fits determined by our SED fitting code with \pegase\ spectra. The tabulated flux percentages for the two-component models in the middle section of the table are the relative contributions from the (young,old) components to the total flux measured in that band.}
    \label{tab:sedres}
\end{table*}

\subsection{Possible AGN contributions}
Throughout the SED analyses described above, we assumed only thermal contributions to the heating of the metal-rich gas. However, given recent observations suggesting possible AGN contributions at high redshift (e.g., \citealt{tilvi16,laporte17b,mainali18}), it is natural to consider how our results might change should there be non-thermal radiation components in our four galaxies.

\subsubsection{Influence on the [\OIII] ratio}
\begin{figure}
\center
 \includegraphics[width=\columnwidth]{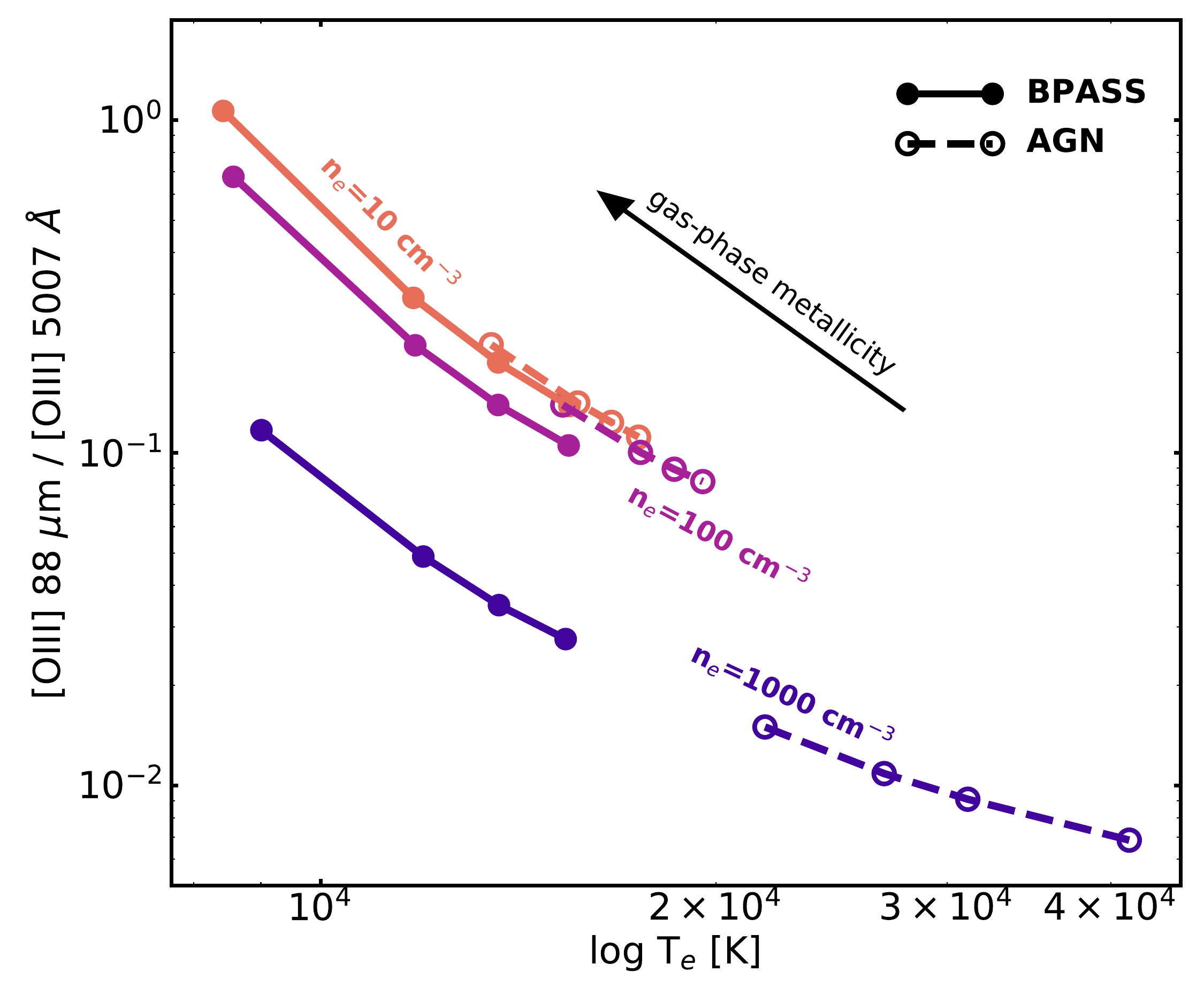}
 \caption{The [\OIII] 88 $\mu$m / [\OIII] 5007 \AA\ line ratio computed over a grid of electron densities and gas-phase metallicities for a stellar radiation field using BPASS (filled points and solid lines) and a AGN (empty points and dashed lines) models using \texttt{Cloudy}. The curves and points are colour-coded according to their constant electron density.
 For clarity the case with $n_e$=3 cm$^{-3}$ is not shown. From left to right, the first, second, third and fourth scatter points in each of the curves represent metallicites of 1, 0.4, 0.2 and 0.05 Z$_{\odot}$. AGN contributions do not significantly alter the allowed [\OIII] line ratio for a given electron density and metallicity, except at very high densities and temperatures.}
 \label{fig:AGN}
\end{figure}

To quantify differences in the FIR/optical [\OIII] ratio arising from excitation by stellar or AGN contributions, we run \texttt{Cloudy} using v2.1 of the Binary Population and Spectral Synthesis (BPASS; \citealt{eldridge17}) models as well as the code's own AGN option (based on the model of \citealt{mathews87}). For both stellar and AGN cases the [\OIII] line ratio is computed over a range of electron densities n$_{e}$=[3,10,100,1000] (cm$^{-3}$) and gas-phase metallicities Z=[0.05,0.2,0.4,1] (Z$_{\odot}$) for a fixed ionisation parameter log $U$=$-$2.5. We show the results of these simulations in Figure \ref{fig:AGN}, which demonstrates the AGN case makes little difference at fixed electron density and gas-phase metallicity (a result, we note, is unaffected by a change in the ionisation parameter). The similarity between the stellar and AGN results is particularly evident for electron densities $n_{e}\lesssim$1000 cm$^{-3}$, where relatively large ratios ([\OIII] 88 $\mu$m / [\OIII] 5007 \AA$\approx$0.1-1) are seen that increase with metallicity. Such a trend is consistent with the results in Section \ref{subsec:almaoiii} and Figure \ref{fig:ratio} and arises because of relatively low electron temperatures ($T_{e}\lesssim$15,000-20,000 $\SI{}{\kelvin}$) where, for low density regimes, the line ratio is influenced primarily by temperature fluctuations rather than by collisional de-excitation. As the electron density increases to $n_{e}\gtrsim$1000 cm$^{-3}$, however, order-of-magnitude differences appear: while both the stellar and AGN cases probe line ratios $<$0.1 favouring particularly strong [\OIII] 5007 \AA\ emission, collisional de-excitation of the [\OIII] 88 $\mu$m line can occur and, combined with the higher temperatures ($T_{e}>$20,000 $\SI{}{\kelvin}$) possible in the AGN case, this increases excitation of [\OIII] 5007 \AA\ whereas [\OIII] 88 $\mu$m is quenched. As a result the ratio is driven down to low values of $\sim$0.01. For our sample of four sources, the measured electron densities and temperatures (see Section \ref{subsec:almaoiii}) probe regions of the parameter space where AGN do not have a significant effect and thus our [\OIII] line ratios are unlikely to be strong affected by non-thermal radiation.

\subsubsection{Observational evidence for AGN at $z>7$?}
Although we have found that the [\OIII] line ratio is only affected by a AGN contribution in regions of very high temperature and electron densities unlikely to be representative of the general $z>7$ population, it is nonetheless informative to consider whether our sample of four galaxies and those in the extended list of IRAC-excess sources in Table \ref{tab:z7} show any observational evidence for AGN activity, and if so at what level. At $z>$7, the most useful diagnostics for distinguishing between thermal and non-thermal contributions are rest-frame UV emission lines and their ratios, in particular \NV\ $\lambdaup$1240,1243 \AA, \CIV\ $\lambdaup$1549 \AA, \HeII\ $\lambdaup$1640 \AA\ and \CIII]\ $\lambdaup\lambdaup$1907,1909 \AA\ \citep{feltre16,laporte17b,stark17,mainali18}. With an ionising energy of nearly 80 eV, the detection of \NV\ alone is a strong indicator of non-thermal radiation \citep{feltre16}. Each of our four \textit{Spitzer}/IRAC- and ALMA-selected objects have been targeted with NIR spectroscopy where the aforementioned emission lines reside at $z>$7. 7.5-10 hrs of VLT/X-Shooter observations revealed only Ly$\alpha$ in JD1 \citep{hashimoto18} and YD4 \citep{laporte17} and no rest-frame UV lines were seen in Y1 \citep{tamura19}. Furthermore, 4 hrs of Subaru/FOCAS observations revealed only Ly$\alpha$ in B14 \citep{furusawa16}. Of the remaining sources compiled in Table \ref{tab:z7}, all but GN-z10-3, MACS1423-z7p64 and GN-108036 have relevant NIR observations, albeit at varying depths. Only four of these reveal one or more detections of the relevant lines: EGSY8p7 (\NV\ $\lambdaup$1243 \AA; \citealt{mainali18}), EGS-zs8-1 (both components of the \CIII]\ $\lambdaup\lambdaup$1907,1909 \AA\ doublet; \citealt{stark17}), z8\_GND\_5296 (\CIII]\ $\lambdaup$1907 \AA; \citealt{hutchison19}) and COSY (\NV\ $\lambdaup$1240 \AA\ and \HeII\ $\lambdaup$1640 \AA; \citealt{laporte17b}). Given their large ionisation potentials, the detection of any of these lines is indicative of extreme radiation fields \citep{feltre16,stark17}. However, to place quantitative constraints on the powering mechanism, accurate measures of both \CIII]\ $\lambdaup$1909 \AA\ and \HeII\ $\lambdaup$1640 \AA\ lines are necessary. As shown in Figure 7 of \citet{laporte17b}, while the ratio \NV\ $\lambdaup$1240 \AA/\HeII\ $\lambdaup$1640 \AA\ constrains the strength of the radiation field, only the \CIII]\ $\lambdaup$1909 \AA/\HeII\ $\lambdaup$1640 \AA\ ratio can robustly distinguish a metal-poor stellar radiation field from one involving an AGN. With the available spectroscopic data, even upper flux limits on these lines do not usefully constrain the nature of the radiation field. Thus, while it is not yet possible to rule out a modest non-thermal contribution, specifically for the four ALMA sources under consideration here, given the absence of any detected high ionisation emission lines, it seems reasonable to conclude any AGN contribution cannot be dominant.

\section{Discussion}
We have shown that a Balmer break, arising from a mature stellar population, may be a significant contributor to the IRAC excess seen in spectroscopically-confirmed $z>7$ star-forming galaxies. While our analysis does not rule out the possibility that much of this excess arises, as has been conventionally assumed, from intense [\OIII] emission, using ALMA [\OIII] $\lambdaup$88 $\mu$m emission and dust mass measures, we have examined whether we can constrain the relative contributions of starlight and line emission.

The distinction between intense line emission, attributed to recent episodes of star formation from a young ($\leqslant$10 Myr) stellar population, and a prominent Balmer break consistent with more mature stars, is important in considerations of the early assembly history of galaxies. Both the stellar masses and earlier star formation histories will differ depending on the relative contributions and this, in turn, will affect the inferred star formation activity beyond the current \textit{HST} redshift horizon of $z\simeq$10. This was first demonstrated for the $z=9.11$ galaxy JD1 by \citet{hashimoto18} where the IRAC excess must arise primarily from starlight, leading to a stellar mass of 4.2$\pm$1.0$\times$10$^9$ M$_{\odot}$ (lens-corrected for the preferred gravitational magnification) only $\simeq$550 Myr after the Big Bang with an implied epoch of first star formation as early as $z\sim$15.

\begin{figure*}
\center
 \includegraphics[width=1.5\columnwidth]{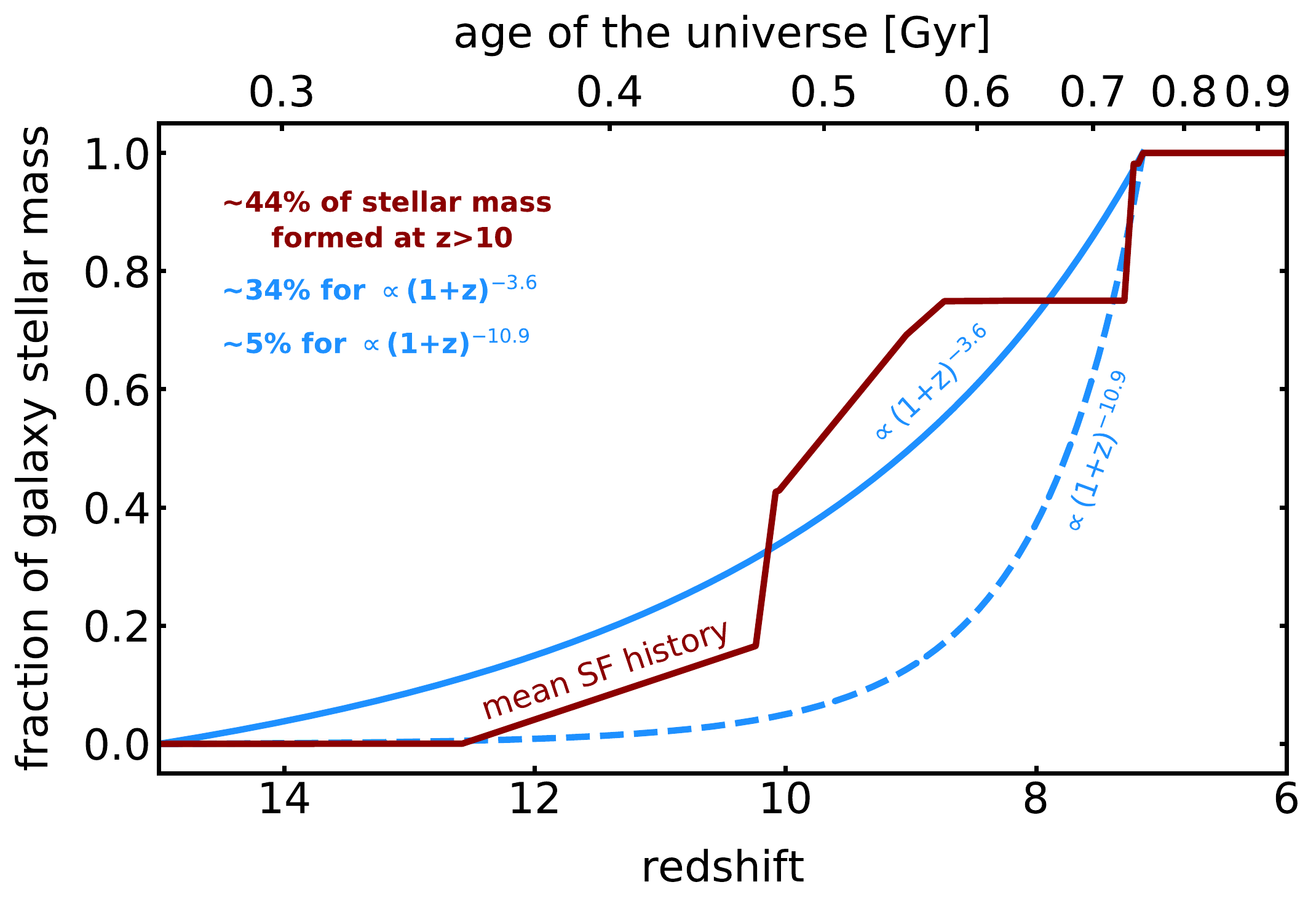}
 \caption{The fractional stellar mass assembly history averaged over the four galaxies in Table \ref{tab:z7oiii} (dark red), adopting the best two-component fits in Figure \ref{fig:sedresults}. The solid and dashed blue lines represent the equivalent fractional histories for the two cosmic SFR density relations presented in Figure 9 of \citet{oesch14}.}
 \label{fig:cd}
\end{figure*}

As an illustration, if we adopt the significant contribution to the IRAC excess from starlight for those $z>7$ sources in Table \ref{tab:z7oiii} for which we fit two-components, their stellar masses increase by an average factor of $\simeq$30 compared to contributions from young starlight alone. Although a single-component fit is probably an extreme comparison in this context, nonetheless our two-component fits with an earlier period of star formation must imply an assembly history beyond $z\simeq$10, as discussed by \citet{hashimoto18}, with interesting consequences for the interpretation of 21cm experiments \citep{bowman18} and the timing of ``cosmic dawn''. In Figure \ref{fig:cd}, we plot the fractional stellar mass assembly history, averaged over our 4 galaxies, up to the epoch of observation in which it can be seen that $\sim$44\% of the stellar mass was produced before a redshift $z\simeq$10. Although clearly a modest sample restricted largely to the brightest studied sources at $z>7$, we can compare this fractional mass assembly history with the prediction of two star formation histories for galaxies with SFRs $>$0.7\,M$_{\odot}$yr$^{-1}$ discussed by \citet{oesch14} similarly normalised at the mean redshift of our galaxies. Whilst the uncertainties are large due to small number statistics, if our galaxies are representative this would indicate a more gradual decline in the star formation history beyond $z\simeq$8 than \citet{oesch14} prefer (see \citealt{mcleod16}). Whilst clearly a simplistic comparison, it serves to emphasise the importance of determining the true origin of the IRAC excess in $z>7$ galaxies. Ultimately NIRSpec on JWST will be well-placed to resolve the ambiguities explored in this paper via direct spectroscopy of a large sample of $7<z<9$ galaxies securing not only the strength of rest-frame optical lines such as [\OIII] $\lambdaup$5007 \AA\ but also absorption line measures such as H$\delta$ which is a further indicator of stellar ages. 

\section*{Acknowledgements}
We thank Rychard Bouwens, Ivo Labb\'e and Dan Stark for useful discussions regarding the interpretation of \textit{Spitzer}/IRAC colours. We would also like to thank Michel Fioc and Brigitte Rocca-Volmerange for their help in generating \pegase\ models, as well as Isabella Lamperti for illuminating discussions on the development of the SED-fitting code. GRB also adds many thanks to Yuichi Harikane and Michael Topping for helping to generate the \texttt{Cloudy} models used in this paper and Roger Wesson for valuable conversations regarding the [\OIII] line ratio. The authors acknowledge funding from the European Research Council (ERC) under the European Union Horizon 2020 research and innovation programme (grant agreement No 669253). NL acknowledges support from the Kavli foundation. 

\section*{Data availability}
The data presented in this paper is available upon email request to the first author.




\bsp  
\label{lastpage}
\end{document}